\documentstyle[fleqn]{book}
\begin{document}

\pagestyle{myheadings}
\markboth{\qquad Lev Vaidman 
\hspace*{\fill}}{\hspace*{\fill}
 Emergence of Weak Values \qquad}

\thispagestyle{empty}
\vspace*{1mm}
{\Large{\bf  Emergence of Weak Values }}

\bigskip by {\large{\sl  Lev Vaidman}\begin{center}
School of Physics and Astronomy,\\
Raymond and Beverly Sackler Faculty of Exact Sciences, \\
Tel-Aviv University, Tel-Aviv 69978, Israel
 \end{center}

\begin{quote}{\bf Abstract. }
Various quantum measurement procedures are analyzed and it is shown
that under certain conditions they yield consistently {\em weak
values} which might be very different from the eigenvalues, the
allowed outcomes according to the standard quantum formalism. The
weak value outcomes result from peculiar quantum interference of the
pointer variable of the measuring device.
\end{quote}

\vskip .3cm
\section{Introduction}
\label{int}
 
In the standard formalism of quantum theory the outcome of a   (good)
measurement must be an eigenvalue of the operator corresponding to the
measured variable. In this paper I will discuss a modified measuring
procedures which will yield instead of an eigenvalue a  {\it weak
  value}, recently
introduced by Aharonov, Albert and Vaidman (1988).  The weak
  value of an observable $A$ is defined for a {\it two-state vector}
  $\langle \Psi_2| ~ |\Psi_1\rangle$ as
\begin{equation}
  \label{weakvalue}
 A_w \equiv {{\langle \Psi_2 | A | \Psi_1 \rangle}
\over {\langle \Psi_2 |\Psi_1 \rangle}} ~~~~.\
\end{equation}
The expectation value of $A$ for a system in a state $|\Psi\rangle$
is a particular case of a weak value when  $|\Psi_1\rangle =
|\Psi_2\rangle =|\Psi\rangle$.

The standard ideal measurements requires infinitely strong
coupling. The weak values emerge only if  the measuring coupling is
bounded  and in most  (but not in all) cases the coupling must be weak
and this is the reason for the name ``weak value''.

The important surprising feature of the weak value is that it might be far away
from the range of the eigenvalues, for example, the weak value of kinetic
energy might be negative, see Aharonov et al. (1993).
The weak value is, in general, a complex number. The (almost) standard
measurement procedure with a weakened coupling yields the real part of
the weak value. The imaginary part can be measured too but we will not
discuss it here.

The expectation value,  $\langle \Psi |A| \Psi \rangle$  emerges in
a {\em weak}  measurement of a
quantum system pre-selected in a state  $|\Psi\rangle$ as well as
in a {\it protective} measurement (Aharonov and
Vaidman 1993, Aharonov, Anandan and Vaidman 1993) when the state
$|\Psi\rangle$ is protected. The weak value (1) emerges in a weak
measurement performed on a quantum system  pre-selected in the state  $|\Psi_1\rangle$ and post-selected
 in the state  $|\Psi_2\rangle$ as well as in a protective measurement
 when the  two-state vector
$\langle \Psi_2| ~ |\Psi_1\rangle$ is protected.
Protective measurements consist of protection coupling and
measuring coupling. The protection coupling usually protects
several quantum states or several quantum two-state vectors. If the system
is protected by such a coupling but not selected in one of the
protected states (two-state vectors) then the outcome of the
measurement is the weak value corresponding to one of the protected
states (two-state vectors) chosen at random. I shall discuss all these
cases below.

\section{Measurement Procedure}
\label{2}

According to  standard
definition, a quantum measurement 
 of a physical variable $A$ is described by the
Hamiltonian: 
\begin{equation}
  \label{ham}
 H = g(t) P A~~~~, 
\end{equation}
where $P$ is a canonical momentum conjugate to the pointer variable  $Q$ of
the measuring  device.
 The
function  $g(t)$ is nonzero only for a very short time
interval corresponding to the measurement, and is normalized so that
$\int g(t)dt=1$.
During  the time of  this 
impulsive measurement, the Hamiltonian (2) dominates the evolution of  the measured
system and the measuring device. Since $[A , H] =0$, the variable $A$ does
not change during the measuring interaction. The initial state of the
pointer variable is usually modeled by a  Gaussian centered at
zero:
\begin{equation}
  \label{phi-in}
\Phi _{in} (Q) = e^{ -{{Q ^2} /{2\Delta ^2}}}.
\end{equation}
Here and below we omit the normalization factor. The pointer is in the ``zero'' position before the measurement, i.e.
its initial probability distribution is
\begin{equation}
  \label{prob}
prob(Q) =  e^{ -{{Q ^2} /{\Delta ^2}}}.  
\end{equation}
 If the initial state of the system is an eigenstate 
$
|\Psi_1 \rangle = |a_i \rangle
$,
then after the interaction (2), the state of the system and the measuring device is:
\begin{equation}
  \label{mdstate}
 |a_i \rangle e^{ -{{(Q-a_i)^2} /{2\Delta ^2}}}.  
\end{equation}
The probability distribution of the pointer variable, $ e^{ -{{(Q-a_i)^2} /{\Delta ^2}}}$ remained unchanged in
its shape, but it is shifted by the eigenvalue $a_i$.
In an ideal measurement, the initial probability distribution of the
pointer is well localized around zero, and thus the final distribution
is well localized around the eigenvalue. Thus, the reading of the
pointer variable in the end of the measurement almost always yields  a
value of the shift (the eigenvalue of the variable).

If the initial state of the system is a superposition 
$
|\Psi_1 \rangle = \Sigma \alpha_i |a_i \rangle
$,
then after the interaction (2) the state of the system and the measuring device is:
\begin{equation}
  \label{state}
 \Sigma \alpha_i |a_i \rangle e^{ -{{(Q-a_i)^2} /{2\Delta ^2}}}.  
\end{equation}
The probability distribution of the pointer variable corresponding to the
state (6) is 
\begin{equation}
  \label{prob2}
prob(Q) = \Sigma |\alpha_i|^2  e^{ -{{(Q-a_i) ^2} /{\Delta ^2}}}.  
\end{equation}
In case of ideal measurement this is a weighted sum of the initial
probability distribution localized around various eigenvalues.
Therefore,  the reading of the
pointer variable in the end of the measurement almost always yields  the
value close to one of the eigenvalues.

In the case of the ideal measurement  the
measuring interaction leads to a very large uncertain change of the system
due to a large uncertainty of the variable $P$. Indeed, in the standard measurement
we require that the pointer shows zero before the measurement, i.e.,
$\Delta$ is very small for
the initial state of the measuring device (3). This requires  large uncertainty in $P$, and therefore the
Hamiltonian (2) causes a large uncertain change.

The weak measurement is also described by the interaction Hamiltonian
(2) but it kept small by taking  
the initial state of the  measuring device  such that   $\langle P \rangle = 0$ and
the uncertainty in $P$ is small.
We consider   $\Delta \gg a_i$ for all  eigenvalues
$a_i$. Then,  we can  perform the Taylor expansion of the sum (7) around
$Q=0$ up to
the first order and rewrite the probability distribution of the
pointer in the following way:
\begin{eqnarray}
  \label{prob2}
 prob(Q) = \Sigma |\alpha_i|^2 e^{ -{{(Q-a_i)^2}
/{\Delta ^2}}} =~~~~~~~~~~~~~~~~~~~~~~~~~~~~~~~~~\nonumber\\
 \Sigma |\alpha_i|^2 (1
-{{(Q-a_i) ^2} /{\Delta ^2}}) = e^{
 -{{(Q-\Sigma|\alpha_i|^2a_i) ^2} /{\Delta ^2}}} . 
\end{eqnarray}
But this is exactly the initial distribution shifted by the value
$\Sigma|\alpha_i|^2a_i$.
 This is the  the expectation value which is  also the weak value in
 this pre-selection case: $A_w =
\Sigma|\alpha_i|^2a_i =\langle \Psi |A|\Psi\rangle$. This weak value
can be found  from statistical analysis of the readings of the measuring
devices of such measurements performed on an ensemble of identical quantum
systems. But it is different conceptually from the standard definition
of expectation value which is a mathematical concept defined from the
statistical analysis of the {\em ideal} measurements of the variable
$A$ all of which yield one of the eigenvalues $a_i$.

\section{Protective Measurements}
\label{3}
 
In general, the weak (expectation) value cannot be measured on a
single system. However, it can be done if the quantum state is {\em
  protected} (Aharonov and
Vaidman 1993). The appropriate measurement interaction
is again described  the Hamiltonian (2), but instead of impulsive
interaction the adiabatic limit of slow and weak interaction is
considered:
$g(t) = 1/T$ for most of the interaction time $T$ and
 $g(t)$ goes to zero gradually before and after the period
$T$.

In this case the interaction Hamiltonian (2) does not dominate the
time evolution during the measurement, moreover, it can be considered
as a perturbation. The free Hamiltonian $H_0$ dominates the evolution. In order to protect a  quantum
state this Hamiltonian must have the state to be a nondegenerate
energy eigenstate.
For $g(t)$ smooth enough we then obtain an adiabatic process in
which the system cannot make a transition from one energy eigenstate
to another, and, in the limit $T \rightarrow \infty$, the interaction
Hamiltonian          changes
 the energy eigenstate by an infinitesimal amount.
   If the initial state
of the system is an eigenstate $\vert E_i\rangle$ 
of $H_0$ then for any given value of
$P$, the energy of the eigenstate shifts by an infinitesimal amount
given by the first order perturbation theory:
\begin{equation}
  \label{sigma}
\delta E = \langle E_i \vert H_{int} \vert E_i  \rangle  =
\langle E_i \vert
A \vert E_i\rangle P/ T.
\end{equation}
 The corresponding time evolution $ e^{-i P \langle E_i \vert
A \vert E_i\rangle } $ shifts the
pointer by the expectation  value of 
$ A$ in the state $\vert E_i\rangle$. 
Thus, the probability distribution of the pointer variable remains unchanged in
its shape, and  is shifted by the
 expectation value 
$\langle A \rangle_i=\langle E_i\vert A\vert E_i\rangle$.

If the initial state of the system is a superposition of several
nondegenerate energy eigenstates 
$
|\Psi_1 \rangle = \Sigma \alpha_i |E_i \rangle
$,
then
a particular outcome $\langle A \rangle_i \equiv \langle E_i \vert A \vert E_i\rangle$ 
appears at random, with the probability $|\alpha_i|^2$.
(Subsequent adiabatic measurements of the same observable $A$ invariably yield 
the expectation value in the same eigenstate $\vert E_i\rangle$.)

\section{ Pre- and Post-Selected Systems}
\label{4}

  Aharonov, Bergmann and
Lebowitz (1964) considered measurements performed on a quantum system between two
other measurements, results of which were given. They proposed describing
the quantum system between two measurements by using two states: the usual
one, evolving towards the future from the time of the first measurement,
and a second state evolving backwards in time, from the time of the second
measurement.  If a system has been prepared at time $t_1$ in a state
$|\Psi_1\rangle$ and is found at time $t_2$ in a state $|\Psi_2\rangle$,
then at time $t$, $t_1<t<t_2$, the system is described by
$
\langle \Psi_2 | e^{i\int_{t_2}^{t} H dt}
{\rm ~~~ and~~~}
e^{-i\int_{t_1}^{t} H dt} |\Psi_1\rangle . 
$
 For simplicity,  we shall consider the free Hamiltonian to
be zero; then, the system at time $t$ is described by the two states 
 $
\langle \Psi_2 |
$ and $ |\Psi_1\rangle $. In order to obtain such a system, we prepare an
ensemble of systems in the state $ |\Psi_1\rangle$, perform a measurement
of the desired variable using separate measuring devices for each system in
the ensemble, and perform the post-selection measurement. If the outcome of
the post-selection was not the desired result, we discard the system and
the corresponding measuring device. We look only at measuring devices
corresponding to the systems post-selected in the state $\langle \Psi_2 |$.

 Let us show briefly how weak values emerge from a measuring procedure
 performed on a pre- and post-selected system
with a sufficiently weak coupling.  We consider a sequence of measurements:
a pre-selection of $|\Psi_{1} \rangle$, a (weak) measurement interaction of
the form of Eq. (2), and a post-selection measurement finding the state
$|\Psi_2 \rangle$.  The state of the measuring device (which was initially
in a Gaussian state) after this sequence is given (up to normalization) by
\begin{equation}
  \label{mdstate}
\Phi (Q) = \langle \Psi_2 \vert
e^{-iPA}
\vert \Psi_1 \rangle e^{ -{{Q ^2} /{2\Delta ^2}}} ~~~~.
\end{equation}
 In the
$P$-representation we can rewrite it  as
\begin{eqnarray}
  \label{sttes}
\tilde \Phi (P) =   \langle \Psi_2
\vert \Psi_1 \rangle ~ e^{-i {A_w} P} ~
e^{-{{\Delta}^2 {P^2}} /{2}} ~ + ~~~~~~~~~~~~~~~~~~~~~~~~~~~~\nonumber\\
\langle \Psi_2 \vert
 \Psi_1 \rangle  \sum_{n=2}^\infty {{(iP)^n}\over{n!}}
[(A^n)_w - (A_w)^n]   e^{ -{{\Delta ^2 P^2}} /{2}}~.
\end{eqnarray}
\noindent
If $\Delta$ is sufficiently large,  we
can neglect the second term of (11) when we Fourier transform
 back to the  $Q$-representation.  Large $\Delta$
corresponds to weak measurement in the sense that the
interaction Hamiltonian
(2) is small.  Thus, in the limit of weak measurement, the final state
of the measuring device (in the $Q$-representation) is
\begin{equation}
  \label{mdstateafter}
\Phi  (Q) = e^{ -{{(Q - A_w)^2} /{2\Delta
^2}}}~~~~. 
\end{equation}
This state represents a measuring device pointing to the weak value,
$A_w$.  Since $\Delta$ has to be large, the weak coupling between a
single system and the measuring device will not, in most cases, lead
to a distinguishable shift of the pointer variable, but collecting the
results of measurements on an ensemble of pre- and post-selected
systems will yield the weak values of a measured variable to any
desired precision. Although we have showed the emergence of weak values
in weak measurements for a specific von Neumann model of measurements,
the result is completely general: any coupling of a pre- and
post-selected system to a variable $A$, provided the coupling is
sufficiently weak, results in effective coupling to $A_w$.

\section{Protection of a Two-State Vector}
\label{5}

 At first sight, it seems that protection of a two-state
vector is impossible. Indeed, if we add a potential that makes one
state  a nondegenerate eigenstate, then the other state, if it is
different, cannot be an eigenstate too. (The states of the two-state
vector cannot be orthogonal.)  But, nevertheless, protection of the
two-state vector is possible (Aharonov and Vaidman, 1995).

 The procedure for protection of a two-state vector of a given system is
accomplished by coupling the system to another  pre- and post-selected
system. The protection procedure takes advantage of the fact that 
weak values  might  acquire complex values. Thus, the effective
Hamiltonian of the protection might not be hermitian. Non-hermitian
Hamiltonians act  in different ways on quantum states evolving forward and
backwards in time. This allows simultaneous protection of two different
states (evolving in opposite time directions).

Let us consider  an example of a  two-state
vector of a spin-1/2 particle, $\langle{\uparrow_y}
|  |{\uparrow_x} \rangle $. The protection procedure uses an external   pre- and post-selected
 system $S$ of a large  spin $N$ that is coupled to our spin via the interaction:
\begin{equation}
  \label{hprot}
H_{prot} = - \lambda {\bf S \cdot \sigma}.
\end{equation}
The external system is pre-selected in the state $|S_x {=} N\rangle$ and
post-selected in the state $\langle S_y {=} N|$, that is, it is described by
the two-state
vector $\langle S_y {=} N
|  |S_x {=} N \rangle $. The coupling constant $\lambda$ is chosen in such a
way that the
interaction with our spin-1/2 particle  cannot
change significantly the two-state vector of the protective system $S$, and
the spin-1/2 particle ``feels'' the effective Hamiltonian in which {\bf $S$} is
replaced by its weak value,
\begin{equation}
  \label{sweak}
{\bf S}_w = {{\langle S_y = N
|(S_x, S_y, S_z)  |S_x = N  \rangle} \over{\langle S_y = N
 |S_x = N  \rangle}} = (N, N, iN)  .
\end{equation}
Thus, the effective protective Hamiltonian is:
\begin{equation}
  \label{heff}
H_{eff} =- \lambda N( \sigma_x +  \sigma_y + i\sigma_z).  
\end{equation}
The state $|{\uparrow_x} \rangle $ is an eigenstates of this
(non-hermitian) Hamiltonian (with eigenvalue $-\lambda N$).  For
backward evolving states the effective Hamiltonian is the hermitian
conjugate of (15) and it has different (nondegenerate) eigenstate with
this eigenvalue; the eigenstate is $\langle {\uparrow_y}|$.

In order to prove  that the Hamiltonian (13) indeed provides the
protection, we have to show that the two-state vector $\langle{\uparrow_y}
|   |{\uparrow_x} \rangle $ will remain essentially unchanged during the
measurement. See details of the proof in  Aharonov and Vaidman, (1995,
1996) and Aharonov et al. (1996).

At least formally we can generalize this method to make a protective
measurement of an arbitrary two-state vector $\langle \Psi _2|  | \Psi
_1 \rangle $ of an arbitrary system.
However, this scheme usually leads to unphysical interaction and is
good only as a gedanken experiment in the framework of non-relativistic
quantum theory where we assume that any hermitian Hamiltonian is possible.

\section{ Weak Values and Protective Measurements}
\label{6}

 The protective Hamiltonian (13) has more
interesting features than just protecting the two state vector 
$\langle{\uparrow_y} | |{\uparrow_x} \rangle$. There is another
two-state vector which is protected: the two state
$\langle{\downarrow_x} | |{\downarrow_y} \rangle$ with corresponding
eigenvalue $\lambda N$.

In general, a nondegenerate non-hermitian Hamiltonian yields protection
for a set of pairs consisting from ``bras''and ``kets''. The
Hamiltonian   can be written in
the following form
\begin{equation}
  \label{non-herm}
H= \Sigma_i \omega_i {{|\Phi_i\rangle \langle \Psi_i |}\over {\langle
\Psi_i| \Phi_i\rangle}} ,
\end{equation}
where $\langle \Psi_i|$ are the ``eigen-bras'' of $H$, and
$|\Phi_i\rangle$ are the ``eigen-kets'' of $H$. The $\langle
\Psi_i|$ form a complete but, in general, non-orthogonal set, and so do
the $|\Phi_i\rangle$. They obey mutual orthogonality condition:
\begin{equation}
  \label{orth}
\langle \Psi_i|\Phi_i\rangle  = \delta_{ij} \langle
\Psi_i|\Phi_i\rangle. 
\end{equation}

 If the initial state is 
a superposition of the eigenstates $|\Psi\rangle = \Sigma_i \alpha_i |\Psi_i\rangle$
then its time evolution is  given by 
\begin{equation}
  \label{evolu}
|\Psi (t)\rangle ={\cal N}(t) \Sigma_i \alpha_i e^{-i\omega_i T}| \Psi_i\rangle
\end{equation}
An adiabatic measurement coupling of a variable $A$  performed on such
system leads to the
 state of the system and the measuring device given by  
\begin{equation}
  \label{sta-md}
 \Sigma_i \alpha_i e^{-i\omega_i T}| \Psi_i\rangle \Phi (Q-{{\langle \Phi_i | A | \Psi_i \rangle}
\over {\langle \Phi_i |\Psi_i \rangle}}) .
\end{equation}
The state of the measuring device is then amplified to a macroscopically
distinguishable situation and, according to standard  interpretation,  a
collapse takes place to  the reading of one of the {\it weak values}
of $A$ with the relative probabilities given by $|\alpha_i
e^{-i\omega_i T}|^2$.

In summary, the main
properties of such adiabatic measurements
 are  (Aharonov et al. 1996):

a) The only possible outcomes of the measurement are the 
weak values  $A_w^i$
 corresponding to one of the pairs of states 
 $\langle \psi_i \vert \vert \phi_i \rangle$
associated with the non hermitian Hamiltonian.

b) A particular outcome $A_w^i$ appears at random, with a probability which
depends only on the initial state of the measured system and 
is independent of the details of the measurement.

c) The measurement leads to an effective collapse 
to the two-state vector $\langle \psi_i \vert \vert \phi_i \rangle$
corresponding to the
observed weak value $A_w^i$.
Subsequent adiabatic measurements of the same observable $A$ invariably yield 
the same weak value.

d) Simultaneous measurements of different observables yield the
weak values corresponding to the same two-state vector 
$\langle \psi_i \vert \vert \phi_i \rangle$.

An  effective non-hermitian Hamiltonian can be obtained
in a real laboratory in a natural way when we consider  a decaying system
and we post-select the cases in which it has not decayed during the
period of time $T$ which is  larger than its
characteristic decay time. Kaon decay is  such an example. $|K_L^0\rangle$ and
$|K_S^0\rangle$ are the eigen-kets of the effective Hamiltonian and they have
corresponding eigen-bras  $\langle K_L'^0|$ and
$\langle {K'}_S^0|$ evolving backward in time. Due to the
$CP- violation$ the states  $|{K}_L^0\rangle$ and
$|K_S^0\rangle$   are not orthogonal. However, the mixing is
small:
$|\langle K_S^0|K_L^0\rangle|\ll 1$, and therefore the  
corresponding backward evolving states are almost identical to the
forward evolving states: $|\langle {K'}_S^0|K_S^0\rangle| = |\langle
{K'}_L^0|K_L^0\rangle| = {1\over
{\sqrt {1 -|\langle
K_S^0|K_L^0\rangle|^2}}}$. Thus, it is difficult to expect a large
effect in this system and for a realistic experimental proposal one
should look, probably, for another system.

\section{Conclusions}

We have shown that weak values emerge in procedures which are very
close to the standard quantum measurements. The procedures are: (i)
weak measurement performed on ensemble of pre-selected quantum
systems, (ii) adiabatic measurement on a single system with a
non-degenerate energy spectrum, (iii) weak measurement on pre- and
post-selected ensemble, (iv) adiabatic measurement on a single system
described by a non-hermitian Hamiltonian.  In cases (i-ii) the weak
values are just expectation values but in cases (iii-iv) the weak
values might lie outside the range of eigenvalues.  These results can
be explained as a peculiar interference effect of the pointer variable
of the measuring device (for computer simulation of these interference
effects see Vaidman, 1995 and Unruh, 1995) but they are most naturally
explained in the framework of the two-state vector formalism.

In fact, the measurements discussed above are not just gedanken
experiments. Experiments of type (i) are frequently performed in
laboratories: in many cases the individual measurement can not reach
the required precision and the measured quantity is found from a
measurement on an ensemble of identically prepared systems (but not
all such cases correspond to weak measurements). Some types of elastic
scattering experiments might fall under category (ii). There were
several experiments of the type (iii). The best example, probably, is
photon polarization measurement (Ritchie, 1991). I do not know about
any performed experiment of type (iv). The most promising is a
subclass of such experiments which consist of adiabatic measurements
performed on a decaying system which has not decayed yet.  We do not
know for what decaying system the weak values can emerge in adiabatic
measurements in today's laboratory. We leave it as a challenge to find
such realistic proposals.

This research was supported in part by grant 614/95 of the Basic
Research Foundation (administered by the Israel Academy of Sciences
and Humanities).

\bigskip{\bf References}\frenchspacing

\begin{list}{\null}{\setlength{\itemsep}{0mm}\setlength{\leftmargin}{4mm}}
\item \hspace*{-4mm}
  Aharonov, Y.,  Albert, D., and Vaidman, L. (1988),
How the Result of Measurement of a Component of the Spin of a
Spin-1/2
 Particle Can Turn Out to Be 100.
{\it Phys. Rev. Lett.  60}: 1351.

\item \hspace*{-4mm}
 Aharonov, Y., Anandan, J., and Vaidman, L. (1993)
Meaning of the wave function,
 {\it Phys. Rev. A 47}: 4616.

\item \hspace*{-4mm}
Aharonov, Y.,  Bergmann,  P.G., and  Lebowitz, J.L. (1964),
Time symmetry in the quantum process of measurement,
 {\em Phys. Rev. 134B}: 1410.

\item \hspace*{-4mm}
Aharonov, Y., Massar, S., Tollaksen, J., Popescu, S., and  Vaidman, L.  (1996),
 Adiabatic measurements on decaying systems,
{\it Phys.  Rev.  Lett.}, to be published. 
 
\item \hspace*{-4mm}
Aharonov, Y.,  Popescu, S., Rohrlich, D., and Vaidman, L.  (1993),
 Measurements, errors, and negative kinetic energy, 
 {\em Phys.   Rev. A 48}: 4084.

\item \hspace*{-4mm}
 Aharonov, Y., and Vaidman, L. (1993), Measurement of the Schr\"odinger wave of a single particle,
 {\it Phys. Lett.  A178}: 38.

\item \hspace*{-4mm}
 Aharonov, Y., and Vaidman, L. (1995),
 Protective measurements,
 {\em Ann. NY Acad. Sci. 480}, 361.

\item \hspace*{-4mm}
 Aharonov, Y., and Vaidman, L. (1996),  Protective measurements of two-state vectors,
 in
``Experimental Metaphysics---Quantum Mechanical
   Studies in Honor of Abner Shimony,'' edited by R.S.Cohen, M. Horne,
   and J. Stachel, Kluwer.

\item \hspace*{-4mm}
Ritchie, N.W.M.,  Story, J.G. and  Hulet, R.G. (1991),
Realization of a measurement of a weak value,
{\em Phys. Rev. Lett. 66}: 1107.

\item \hspace*{-4mm}
Vaidman. L., (1995), Weak measurements,
in  {\it Advances in Quantum Phenomena}, E. Beltrametti and J.M.
Levy-Leblond eds.,  NATO ASI Series B: Physics  Vol. 347, Plenum
Press, NY, p. 357.

\item \hspace*{-4mm}
Unruh, W.G. (1995), Varieties of quantum measurements,
  {\it Ann. NY Acad. Sci. 755}, 560.

\end{list}

\end{document}